# Efficacy of Boron Nitride Encapsulation against Plasma-Processing in van der Waals Heterostructures

Pawan Kumar[1,3], Kelotchi S. Figueroa[2], Alexandre C. Foucher[3], Kiyoung Jo[1], Natalia Acero[4], Eric A. Stach[3,5#] and Deep Jariwala[1#]

[1]Department of Electrical and Systems Engineering, University of Pennsylvania, Philadelphia-19104, USA.

[2]Department of Physics and Electronics, University of Puerto Rico, Humacao-00791, Puerto Rico.

[3]Department of Materials Science and Engineering, University of Pennsylvania, Philadelphia-19104, USA.

[4]Vagelos Integrated Program for Energy Research, University of Pennsylvania, Philadelphia-19104, USA.

[5]Laboratory for Research on the Structure of Matter, University of Pennsylvania, Philadelphia, 19104, USA

#E-mail: dmj@seas.upenn.edu; stach@seas.upenn.edu

**ABSTRACT:**

Two-dimensional (2D) transition metal dichalcogenides (TMDCs) are the subject of intense investigation for applications in optics, electronics, catalysis, and energy storage. Their optical and electronic properties can be significantly enhanced when encapsulated in an environment that is free of charge disorder. Because hexagonal boron nitride (h-BN) is atomically thin, highly-crystalline, and is a strong insulator, it is one of the most commonly used 2D materials to encapsulate and passivate TMDCs. In this report, we examine how ultrathin h-BN shields an underlying $MoS_2$ TMDC layer from the energetic argon plasmas that are routinely used during semiconductor device fabrication and post-processing. Aberration-corrected Scanning Transmission Electron Microscopy is used to analyze defect formation in both the h-BN and $MoS_2$ layers, and these observations are correlated with Raman and photoluminescence spectroscopy. Our results highlight that h-BN is an effective barrier for short plasma exposures (< 30 secs) but is ineffective for longer exposures, which result in extensive knock-on damage and amorphization in the underlying $MoS_2$.

## INTRODUCTION:

The isolation of layered van der Waals crystals into atomically-thin, two-dimensional (2D) structures has led to significant new insights in condensed matter physics[1-2], which have in turn led to fundamentally new electronic device designs.[3-4] Significant effort has been spent on the study of synthetic routes, fundamental physical phenomena, and device properties in these systems. However, there has been less focus on device processing. Ultimately the applications of all 2D materials will require precise control over crystalline quality, thickness (layer number), and, thus, over device processing conditions. With this background, we address an old but relevant problem associated with a ubiquitous process in microelectronics: plasma processing.

Plasma processing is widely used to clean, functionalize and passivate surfaces, as well as to etch materials.[5-11] It has been applied to 2D materials since the early days of graphene device research.[12-15] However, it was soon realized that energetic plasmas could affect the structural and chemical stability of 2D materials and thereby degrade lateral transport in electronic devices.[16-18] Therefore high-quality, stable, and scalable encapsulants are needed to protect 2D device channels, and they continue to be an object of research. However, it is unclear if plasma processing leads to charge incorporation in encapsulating layers that degrades carrier transport in a structurally intact 2D material or if it directly damages the material. Often charge transport measurements are used to infer the role of defect formation in these cases, but transport measurements only provide indirect evidence of defect introduction. These facts motivate us to explore how plasma etching conditions affect both the encapsulation layer and the active channel using a direct approach.

In all high-performance 2D semiconductor devices, it is essential to isolate or encapsulate the active semiconductor layers to limit charge inhomogeneities and exposure to processing chemicals. Encapsulation is needed to protect electronic devices such as transistors, where the channel is buried under a dielectric insulator, and optoelectronic devices where the junction is buried under contacts and barrier layers. Charge inhomogeneity results from trapped charges, dangling bonds, and dipoles of ionic bonds, all of which impede electronic transport and inhibit radiative recombination in 2D layers. Organic layers (polymers) and flat, highly-crystalline, and nearly covalent materials such as hexagonal boron nitride (h-BN)[19] have proven to be effective substrates and encapsulants for 2D channels and active layers[20-]

[22]. However, polymers and small organic molecules are susceptible to thermal damage as well as swelling and dissolution upon solvent exposure. This means that they are unsuitable as permanent encapsulants during semiconductor device fabrication and processing. However, h-BN possesses high chemical and thermal stability, making it a potentially superior encapsulant. Several schemes for direct growth and transfer of h-BN encapsulated graphene and 2D semiconductor devices have been developed[23-26]. In these studies, a post-lithography etching step is essential for defining channels and contacts. Although several studies assume that h-BN is adequately protecting the underlying active 2D layer[27-30], there have been no systematic studies that provide mechanistic insight into its effectiveness. In this study, we perform a systematic investigation of the efficacy of the h-BN layer as an encapsulant. We correlate optical spectroscopy and atomic-resolution imaging analysis to understand how plasma dose variation, sequential plasma exposure, and encapsulant and underlayer thickness affect the rates of damage accumulation.

## EXPERIMENTAL SECTION:

**Materials and Methods:**

Mechanically exfoliated 2D $MoS_2$ layers were prepared using the conventional scotch tape method, as described elsewhere[31]. The thickness of the exfoliated layers was intentionally chosen, such that each sample could be reproduced for a number of repetitive analyses. Exfoliated h-BN and $MoS_2$ layers were transferred to the $SiO_2$/Si substrate by dry transfer technique utilizing a poly-dimethyl siloxane (PDMS) stamp. The dry transfer technique uses a motorized micromanipulator stage (X-Y-Z axes), attached with an optical microscope along-with with a home-made heating stage (based on pyroelectric material). After transferring $MoS_2$ layers to $SiO_2$/Si substrates, samples were annealed in a quartz tube furnace in a closed gas (Ar+ $H_2$) environment to remove all PDMS contaminants. Annealing was performed at 300 ˚C for 4 hrs. to clean contaminants as well as release the strain developed during the pressure-based dry transfer method. Similarly, h-BN and $MoS_2$ layers were transferred to a dedicated SiNx TEM grid (Norcada Inc., 3x3 array of 100 μm diameter holes) and annealed in the same manner to remove PDMS contaminations prior to irradiation and subsequent STEM characterization. Two-different plasma irradiation systems are used here for the different samples at different exposure times. Ultra-pure Ar gas (99.995%) was used in all the plasma exposure analysis. A dedicated plasma cleaning system was available for holding TEM holders such that we can treat samples for multiple exposure times while the TEM grid installed into TEM holder. Plasma exposure time for TEM analysis (sample on TEM grid) vs. Raman analysis (sample on $SiO_2$/Si substrate) are different as sample positioning distance from the plasma ignition source is different in both the systems based on the available configuration.

**Characterization:**

Raman spectroscopy as well as Photoluminescence measurement is carried out using LabRAM HR Evolution HORIBA system. A 633 nm LASER with spot size ~0.5 µm, 1% laser power was used for diagnosed defect related analysis in h-BN as well as $MoS_2$ layers. For the case PL analysis, 405 nm laser was used with 0.1 % laser power (0.25 µWatt) with 1 sec acquisition time. Optical microscope from Olympus, USA was used to capture and analyze all the samples before and after plasma treatment. The plasma irradiation system (Tergeo Plasma Cleaner, PIE Scientific) was used with 65 sccm of Ar (99.995% purity) flow and 50-Watt transmitted RF power under 0.35 Torr base vacuum environment. The same parameters were utilized in the case of a dedicated TEM plasma cleaner system (Gatan, Solarus 950). HAADF-STEM has been used to directly visualize all the samples for defects evolution. An aberration-corrected JEOL NEOARM STEM, operating at an accelerating voltage of 200 kV and convergence angle of 25-29 mrad is used for all samples. For the JEOL NEOARM STEM, the condenser lens aperture was 40 µm with a camera length of 4 cm for imaging and the probe current was 120 pA. All of the captured STEM images were collected using Gatan GMS software and associated Gatan bright-field and high angle annular dark field detectors. Experimentally acquired STEM images are smoothed using the adaptive gaussian blur function (with a radius of 1-2 pixels) available in ImageJ.

## RESULTS AND DISCUSSION:

During sample preparation, device fabrication, and post-processing a 2D material or heterostructure is likely to encounter several forms of energetic radiation sources. Radiation sources that are used include electron, ion (Ar, He, Xe), and photon (laser, UV) fluxes. However, ion beams and UV light are the most routinely adopted in etching and lithography processes, respectively. Here, we investigate the etching behavior of h-BN, $MoS_2$, and their heterostructures when subject to $Ar^+$ ion plasma exposure. Argon ion plasmas are the most commonly used as argon does not form chemical bonds with the sample due to its inherent inertness. Argon is also a heavy enough ion that can provide sufficient kinetic energy to etch samples at reasonable accelerating voltages.

**Fig. 1** shows a schematic of the physically stacked 2D heterostructure samples used in this study. Thin flakes of h-BN and $MoS_2$ were mechanically exfoliated from the bulk using Scotch tape. They were subsequently transferred onto oxidized silicon wafers or dry PDMS stamps.

Both heterostructures and samples transferred onto scanning transmission electron microscopy (STEM) heating platforms were created using direct transfer via viscoelastic PDMS stamps. Further details of the sample preparation and transfer methods are provided in the Materials and Methods section.

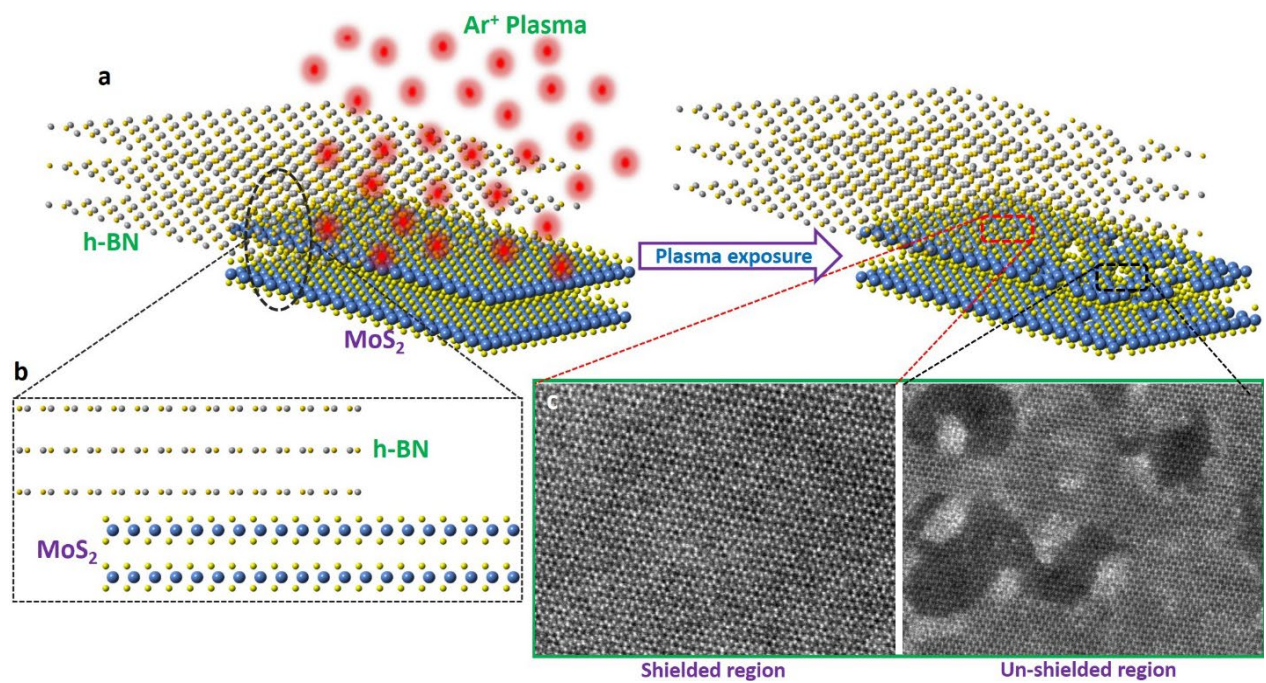

**Fig. 1:** (a) Schematic representing MoS$_2$/h-BN partially overlapping van der Waals heterostructures under investigation in this study before and after Ar$^+$ plasma exposure. (b) Cross-sectional schematic of the heterostructure. (c) Corresponding STEM images show an atomic-scale structural changes across shielded and un-shielded region after the 15 secs exposure of plasma.

Raman spectroscopy is used to provide a global measure of defect formation as a function of plasma exposure for both h-BN shield and un-shielded MoS$_2$ (**Fig. 2**). Fig. 2a presents an optical microscopy (OM) image of a stacked heterostructure of h-BN (circled). There are also regions of pure h-BN and pure MoS$_2$ adjacent (arrows). Height images and corresponding h-BN thickness profiles are presented for the same heterostructure (h-BN/MoS$_2$) in Fig. 2b. Raman spectra as a function of Ar$^+$ exposure time from the marked un-shielded and shielded MoS$_2$ regions shown in Fig. 2c. The distinct peak at 226 cm$^{-1}$ [labeled LA (M)] has been associated in prior work with the appearance of defects arising from the scattering of phonons at the Brillion zone edge.[32-34] Similarly, E$_{2g}$ and A$_{1g}$ mode intensities decrease with increase of plasma exposure time, confirming that the lattice experiences increasing damage with exposure time. Control data from the h-BN layer is presented as Fig. S1 and S2, and changes in optical contrast in the OM image are presented in Fig. S3. Pristine MoS$_2$ has a very small LA

(M) Raman mode as a monolayer and is negligible for few layer MoS$_2$: this LA mode most likely originates from defects created during the exfoliation and transfer process. Two and half minutes of plasma exposure leads to a significantly enhanced LA (M) signal, indicating the formation of a significant quantity of defects. After 5 minutes of plasma exposure, the intensity of the LA (M) Raman mode is of the same magnitude as the lattice $A_{1g}$(M) - LA (M) phonon mode, indicating significant damage accumulation.

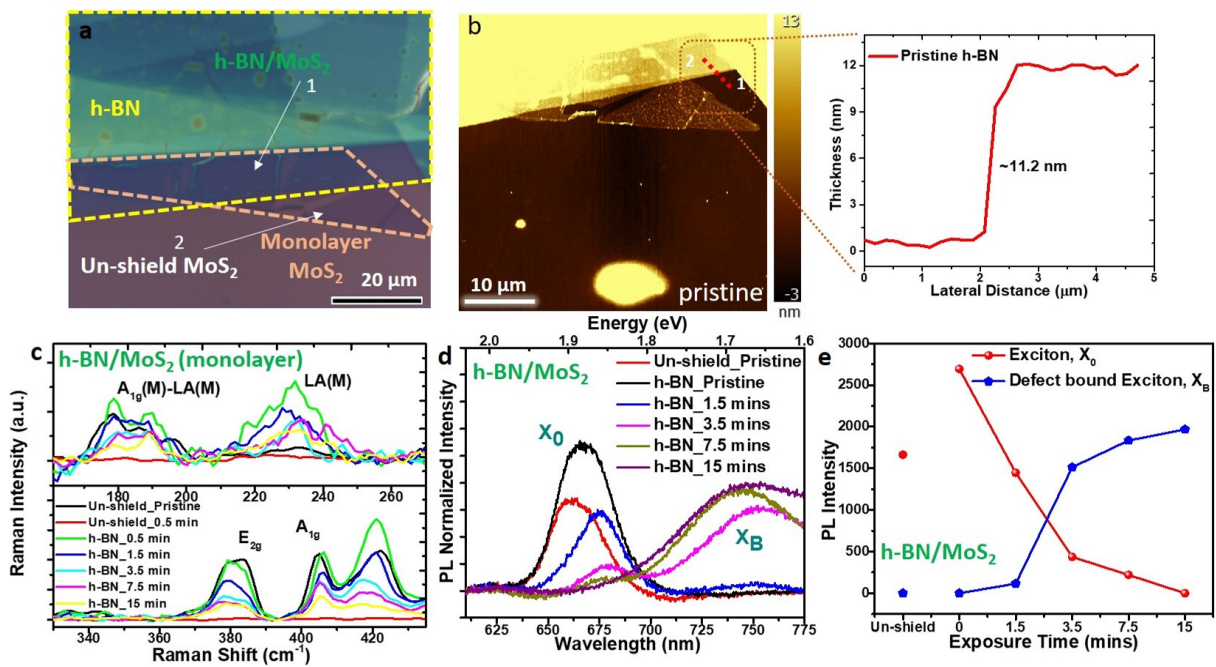

**Fig. 2:** Raman and photoluminescence (PL) characteristics of shielded vs. un-shielded monolayer MoS$_2$ for different Ar$^+$ plasma exposure times. (a) Optical microscope image of few-layer MoS$_2$ shielded by few-layer h-BN, and the corresponding (b) AFM height image of heterostructure region along-with height profile of the h-BN layer (~11.2 nm) in its pristine form. (c) Raman spectra at different plasma exposure for un-shielded and shielded monolayer MoS$_2$. (d) PL response of monolayer MoS$_2$ analyzed for same h-BN shielded monolayer MoS$_2$ region shown in the optical micrograph and corresponding (e) PL peak of $X_0$ and $X_B$ intensity variation with the increasing plasma exposure.

We have also examined the Ar$^+$ plasma exposure effects for shorter intervals of time (within a one-minute duration, using 20 secs step sizes) across different layer thicknesses of un-shielded MoS$_2$ and report those results in supplementary Information Fig. S4. We find that defect formation occurs at levels measurable via Raman Spectroscopy upon Ar$^+$ plasma exposure for time scales as short as 30 secs for an unprotected monolayer MoS$_2$ sample. In contrast, Ar$^+$ plasma exposure times of 2.5 and 5 minutes showed a less significant rise in the LA (M) Raman signal for the region of the MoS$_2$ flake that was shielded under a few-layers of h-BN, as shown in Fig. S5. While Raman spectroscopy is a reliable way to ascertain lattice damage, Raman signals from 2D materials are inherently inefficient. Photoluminescence is a far more sensitive measure of crystal quality and defect density for a high-quality, direct band gap semiconductor. The smallest deviation from perfect crystalline order can induce non-radiative recombination that reduces the primary PL efficiency[35]. Deviations from crystalline order can also introduce trap states that lead to photoluminescence that is red-shifted from the original PL peak location[35-36].

We have recorded the PL response after plasma exposure for shielded and un-shielded monolayer $MoS_2$. Fig. 2a presents an OM image of the heterostructure sample from which we obtained PL spectra. Region 1 has a monolayer of $MoS_2$ that is shielded under a thin h-BN flake (~11.2 nm thin), and region 2 is unshielded monolayer $MoS_2$. The PL data is presented as Fig. 2d-e. The primary peak at ~652 nm (~1.9 eV) is caused by $X_0$-exciton emission,[37] and the peak at ~740 nm (~1.67 eV) is known to be a defect bound exciton peak.[38-39] Similar PL spectra for low incident light intensity have observed previously in the literature for $MoS_2$.[40-42] The PL intensity of the unshielded $MoS_2$ region, before irradiation ("pristine") is shown in red on Fig. 2d. The PL intensity of the un-shielded $MoS_2$ flake is comparatively weaker than the shielded region for pristine samples. This is an optical effect relate to the formation of the heterostructure: the h-BN has a higher index of refraction, while the $SiO_2$ wafer has a lower index of refraction. The higher index medium thus enables increased light extraction. With increasing plasma exposure, we see two effects, which we summarize in Fig. 2e. First, with increasing $Ar^+$ plasma exposure, the peak associated with defect bound exciton emission ($X_B$) increases, clearly indicating the increase in defect content in the layer as the incident energetic $Ar_+$ ions penetrate the h-BN shield. Second, there is a concurrent, correlated decrease in the primary neutral exciton ($X_0$) peak intensity. Interestingly, the h-BN/$MoS_2$ heterostructure shows a sudden increase in the $MoS_2$-$X_B$ intensity with 3.5 min and greater plasma exposure time. This data suggests that the h-BN layer is an effective shield in heterostructure up-to a certain threshold of plasma exposure only.

Optical and vibrational spectroscopy are an effective way to track general trends in damage accumulation. However, they do not provide atomic-scale information about the accumulation of individual defects. To complement these spectroscopies, we have performed extensive characterization using aberration-corrected high-angle annular dark field (HAADF) scanning transmission electron microscopy (STEM) imaging. Flake samples were prepared using the same methodology described in Fig. 1 and transferred onto a SiNx TEM grid that contains separated 100 μm diameter holes. This allowed the heterostructures to be irradiated with Ar$^+$ plasma without damaging the underlying SiNx membrane. Details of the plasma irradiation system used for analyzing all STEM samples are described in the experimental methods section. A basic MoS$_2$ structural model and low magnification as well as atomically resolved HAADF STEM imaging for pristine and different plasma exposed MoS$_2$ regions are shown in Figure S6-S7. HAADF STEM imaging was performed after successive plasma exposures, with varying time intervals (5, 10, and 15 secs). This leads to four sample conditions: "pristine" (0 sec exposure) and 5, 15, and 30 secs of cumulative exposure, respectively, as shown in **Fig. 3**. We compare the defect evolution analysis for two different regions of the same sample, an h-BN shielded MoS$_2$ heterostructure versus an unshielded (bare) MoS$_2$ region. It is worth noting that even though a vertical van der Waals heterostructure of h-BN over MoS$_2$ is being imaged in transmission in Fig. 3a-d, nearly all of the image signal is from the MoS$_2$ layer. There are two reasons for this. First, aberration-corrected STEM images have a very small depth of field, and we maintained the focus on the MoS$_2$ layer.[43] Second, the intensity in HAADF images scales with $Z^{1.65}$,[43] and both Mo and S are significantly heavier than B and N. Nonetheless, A Fast Fourier transformation (FFT) diffractogram can detect the periodicity in the image from both the MoS$_2$ and h-BN lattices, and it indicates that there is a 12.2° twist angle between them. Fig. 3a-c shows that the h-BN shielded MoS$_2$ heterostructure experiences negligible defect formation until a cumulative 15 secs of plasma exposure. However, after 30 cumulative seconds of plasma exposure, the h-BN/MoS$_2$ heterostructure shows visible lattice damage (circled).

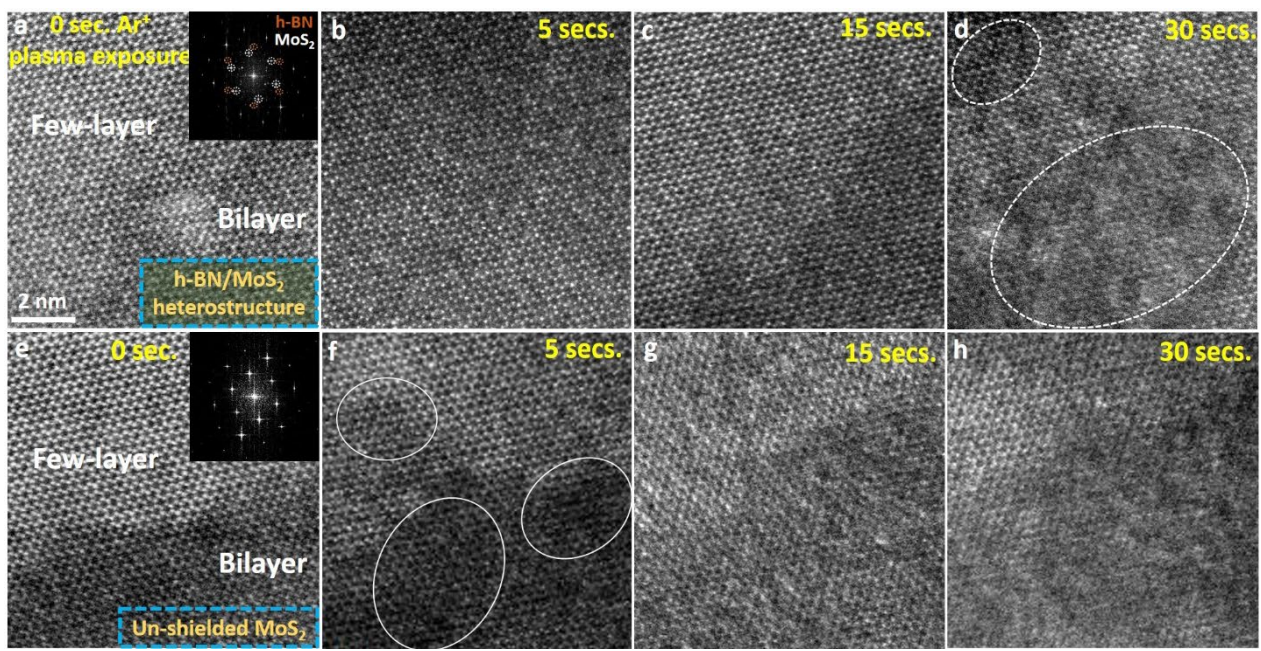

**Fig. 3:** Aberration corrected HAADF-STEM images from pristine and Ar$^+$ plasma exposed samples. (a-d) Images from the h-BN/few-layer MoS$_2$ heterostructure region showing minimal defect/damage creation up to 15 secs (a-c) followed by significant damage at 30 secs (d). In contrast, (e-h) STEM imaging from un-shielded MoS$_2$, shows significant lattice damage starting from 5 secs plasma exposure (f) with near amorphization of bi-layer region by 30 secs exposure to Ar$^+$ plasma (h). Insets (a and e) are FFT diffractograms indicating the presence of h-BN/ MoS$_2$ heterostructure vs. bare un-shielded MoS$_2$, respectively; all images are at the same scale with a representative scale bar shown in a.

In contrast, bare MoS$_2$ shows signatures of visible lattice damage after merely 5 secs of exposure (Fig. 3f). Contrast in the different HAADF STEM images (from pristine to varying plasma exposure time) are not quantitatively comparable since each individual image is self-normalized by the image acquisition software (Gatan GMS) during image acquisition to its maximum intensity. Gray-scale histograms for the images in Fig. 3 are presented in Fig. S8. Each histogram has been normalized to the respective maximum image intensity, and scaled to a common distribution based on the mean intensity and standard deviation. All the image processing and normalization carried out using "Sci-kit" as well as "Fiji" to read and process the images".[44-45] Lattice damage is induced by the plasma in both the bilayer and the few-layer portion of the MoS$_2$ flake following just 5 secs of exposure. Additional plasma exposure (Fig. 3f-h) leads to progressively more severe lattice damage, resulting in near amorphization of the bilayer region after 30 secs of plasma exposure (Fig. 3h). These images also indicate that the damage grows in spatially localized regions with increasing exposure. In other words, damage accumulates at defects introduced at earlier times, not through continued re-nucleation. We summarize that the defective regions created by that plasma are passivated by ambient atmospheric species that accumulate during transfer from the plasma chamber to the TEM column. This chemical passivation stabilizes these regions, and subsequent plasma exposure leads to increased damage at these undercoordinated sits. Additional low magnification STEM images for both the heterostructure and bare sample regions are shown in Fig. S6, and another set of atomic-scale STEM image showing clustered patches of defected areas after 30 secs of plasma exposure are presented in Fig. S9 (supplementary information). We have performed electron energy loss spectroscopy (EELS) measurements after 15 secs of plasma exposure time (supplementary information, Fig. S10). The atomically resolved EELS data showed the presence of adsorbed oxygen, which is found in defective 2D $MoS_2$ basal planes[46]. Mo and S signals are weaker since the thickness of $MoS_2$ layer (~4-5 nm) is very thin in comparison with the total sample thickness which includes a 20 nm SiNx membrane underneath SiNx as well as h-BN layer (~10nm) covering on the top. We have also studied EELS for a heterostructure (h-BN/$MoS_2$) which undergoes continuous plasma exposure up-to 20 secs. We have first acquired the core loss EELS spectra for Oxygen in pristine state and then after 20 secs plasma exposure from the same region for unshielded as well as shielded region of $MoS_2$ flake, as presented in Fig. S11. A slight increased amount of Oxygen has only observed. This observation supports our hypothesis that ambient exposure stabilizes defects during transfer in and out of the microscope.

To understand the effect of air exposure during sequential plasma bombardment, we subjected a different heterostructure stack to a continuous plasma exposure for 15 seconds. We observed widespread damage and etching of the basal plane in the un-shielded $MoS_2$, as shown in **Fig. 4**. This is in stark contrast with the sequential plasma exposure combined ambient air exposure (vide supra, Fig. 3g), where the lattice damage is more uniform. We see that increasing the total plasma exposure creates a large number of defects in the h-BN as well as the un-shielded $MoS_2$ layer, as shown in Fig. 4a and b, respectively. Furthermore, the defects created here are much larger and are not individual point defects. Instead, holes of 3-5 nm diameter form. This suggests that without the stabilization provided by oxygen during the transfer process, there is an accelerated accumulation of defects during irradiation. Following the nucleation of an individual point defect, the under-coordinated atoms can readily be knocked off their lattice sites, which, combined with extended migration during the continuous plasma exposure, could lead to the growth of substantially large voids.

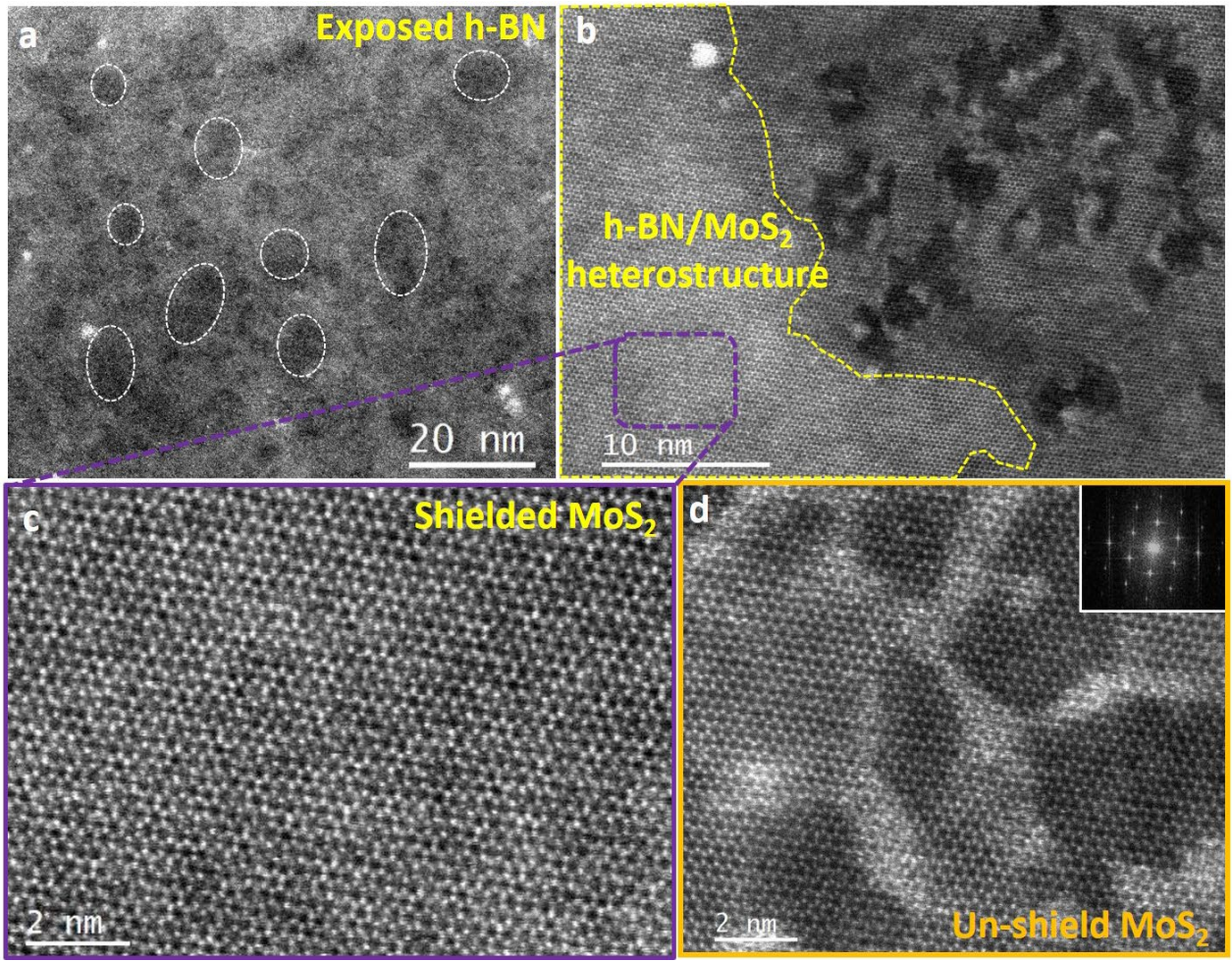

**Fig. 4:** Characterization of a sample region following 15 seconds of Ar$^+$ plasma exposure. (a) Lower magnification HAADF-STEM imaging of a region containing just the h-BN layer. Darker patches of damage are circled. (b) Atomic-scale HAADF STEM image from a sample region containing both h-BN shielded and unshielded MoS$_2$. The region marked in yellow corresponds to the h-BN shielded MoS$_2$ region (left), while the right part of the image is the unshielded region (c) Atomic-scale image of the h-BN shielded MoS$_2$ region shows no strong damage. (d) Atomically resolved STEM image of an unshielded region shows substantial damage.

The h-BN / MoS$_2$ heterostructure remains remarkably intact, as we can see in Fig. 4b and c. This is consistent with the ability of few-layer h-BN to be an effective shield, as shown in the PL data of Fig. 2. We have also performed STEM imaging for a continuous plasma exposure of 20 secs. Figure S12 presents AFM height images and corresponding thickness profiles from the different section of h-BN/MoS$_2$ heterostructures. Furthermore, the same flake was analyzed using Raman spectroscopy and the defect mode is clearly seen in the case of unshielded MoS$_2$ (Figure S13). Afterwards, STEM imaging from the same region of two different sample configurations – unshielded and shielded MoS$_2$ – is presented after 20 secs continuous plasma exposure in Fig. S14. We have also separately examined the defect evolution, and corresponding lattice damage within just the h-BN layers that are being used as a shield in the heterostructure the contrast on bare h-BN in Fig. 4a is not as clear. However, the voids are visible and marked. We have observed similar void formation and damage patches in h-BN as well upon continuous plasma exposure, see supplementary information Fig. S15-S17 for further details. We hypothesize that these localized voids act as a channel for further penetration of plasma ions through layer by layer etching, ultimately reaching the underlying layer, which then causes localized damage in the $MoS_2$ (supplementary information, Fig. S9). Again, this is in stark contrast with sequential exposure to plasma in multiple small steps with ambient air contact between steps. The creation of defective patches and corresponding voids can be easily seen in the AFM height image of h-BN layers (supplementary information, Fig. S18).

## CONCLUSIONS:

We have studied the dynamics of $Ar^+$ plasma-induced defect generation and etching in atomically thin van der Waals heterostructures of h-BN and $MoS_2$. We observe that h-BN effectively shields underlying layers from plasma damage. An atomic-scale imaging suggests that plasma-induced lattice damage is instantaneous for unshielded $MoS_2$, whereas shielded $MoS_2$ is protected by the h-BN up until a certain extent of exposure, and that the extent of damage as function of the time of exposure depends upon the h-BN thickness. Finally, we conclude that continuous plasma exposure is more damaging, as opposed to via sequential exposure. These results indicate that h-BN encapsulation does provide limited protection to underlying $MoS_2$ layers during plasma processing, but that the level of protection varies on a range of parameters.

## ASSOCIATED CONTENT

**Supporting Information**

The Supporting Information is available free of charge at https://pubs.acs.org/doi/xxxxxxx.

**Conflicts of interest:**

There are no conflicts of interest to declare.

## ACKNOWLEDGMENTS:

This work was carried out in part at the Singh Center for Nanotechnology at the University of Pennsylvania, which is supported by the National Science Foundation (NSF) National Nanotechnology Coordinated Infrastructure Program grant NNCI-1542153. DJ, EAS and PK acknowledge primary support for this work from via the NSF DMR Electronic Photonic and Magnetic Materials (EPM) core program grant (DMR-1905853) as well as the University of Pennsylvania Laboratory for Research on the Structure of Matter, a Materials Research Science and Engineering Center (MRSEC) supported by the National Science Foundation (DMR-1720530) . KSF was supported by the LRSM MRSEC REU and the Penn-UPRH Partnership for Research and Education in Materials (PREM), NSF-DMR-1523463 Program. NA and DJ acknowledge support from Vagelos Integrated Program for Energy Research at the University of Pennsylvania as well as the Center for Undergraduate Research and Fellowships at Penn. DJ also acknowledges support for this work by the US Army Research Office under contract number W911NF1910109. ACF and EAS would like to acknowledge the Vagelos Institute for Energy Science and Technology at the University of Pennsylvania for a graduate fellowship to ACF. PK thank James Horwath to help in the STEM image histogram analysis. The authors thank Douglas Yates and Jamie Ford in the Singh Center for Nanotechnology for help with the TEM/STEM measurements.

# Supplementary Information

# Efficacy of Boron Nitride Encapsulation against Plasma-Processing in van der Waals Heterostructures

**Pawan Kumar**[1,3], **Kelotchi S. Figueroa**[2], **Alexandre C. Foucher**[3], **Kiyoung Jo**[1], **Natalia Acero**[4], **Eric A. Stach**[3,5#] **and Deep Jariwala**[1#]

[1]Department of Electrical and Systems Engineering, University of Pennsylvania, Philadelphia-19104, USA.

[2]Department of Physics and Electronics, University of Puerto Rico, Humacao-00791, Puerto Rico.

[3]Department of Materials Science and Engineering, University of Pennsylvania, Philadelphia-19104, USA.

[4]Vagelos Integrated Program for Energy Research, University of Pennsylvania, Philadelphia-19104, USA.

[5]Laboratory for Research on the Structure of Matter, University of Pennsylvania, Philadelphia, 19104, USA

#E-mail: dmj@seas.upenn.edu; stach@seas.upenn.edu

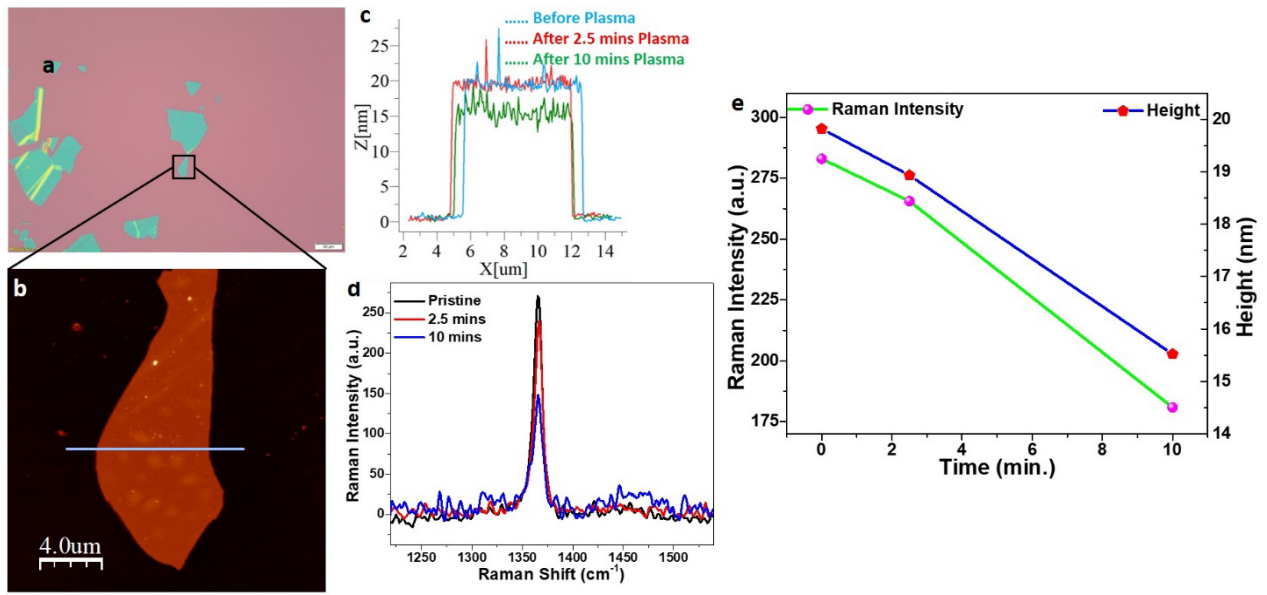

**Fig. S1:** (a) Optical microscopy image of as-transferred and plasma cleaned few-layer h-BN and corresponding (b) AFM image. (c) AFM height profile as a function of indicated plasma exposure time. (d) Raman spectra as a function of indicated plasma exposure time. (e) Correlation of Raman spectra and AFM height as a function of plasma exposure time.

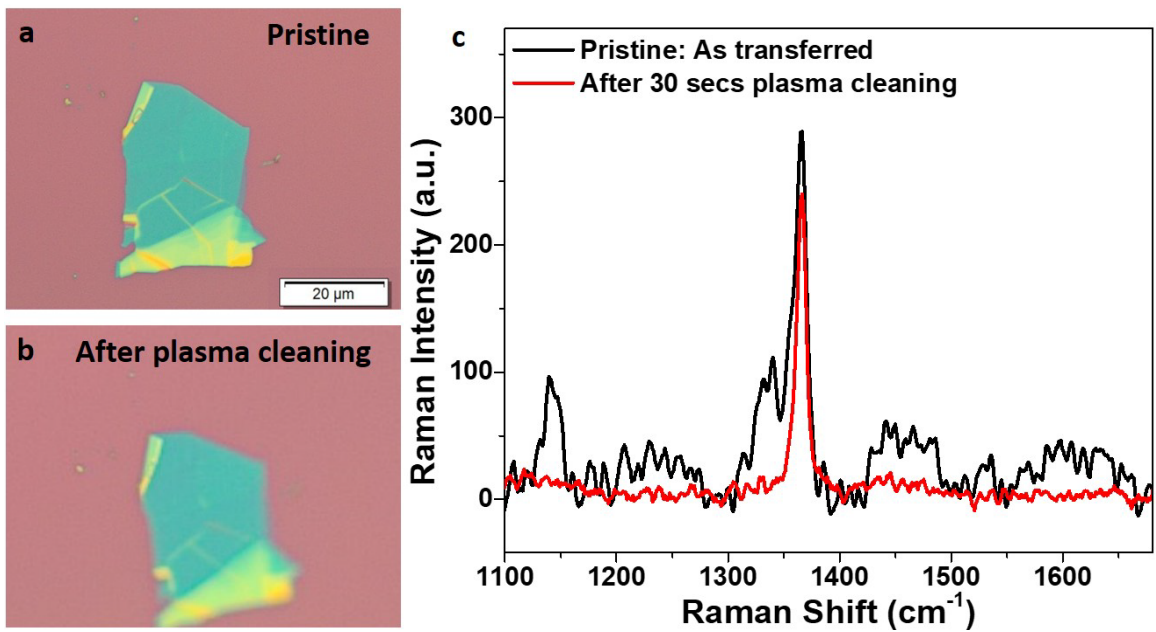

**Fig. S2:** (a-b) Optical microscope image of few-layer h-BN before and after plasma cleaning respectively to remove PDMS contamination and (c) corresponding Raman spectrum which shows the disappearance of peaks associated with PDMS. Only the 1365 cm$^{-1}$ ($E_{2g}$ mode in h-BN) remains after 30 secs of plasma cleaning.

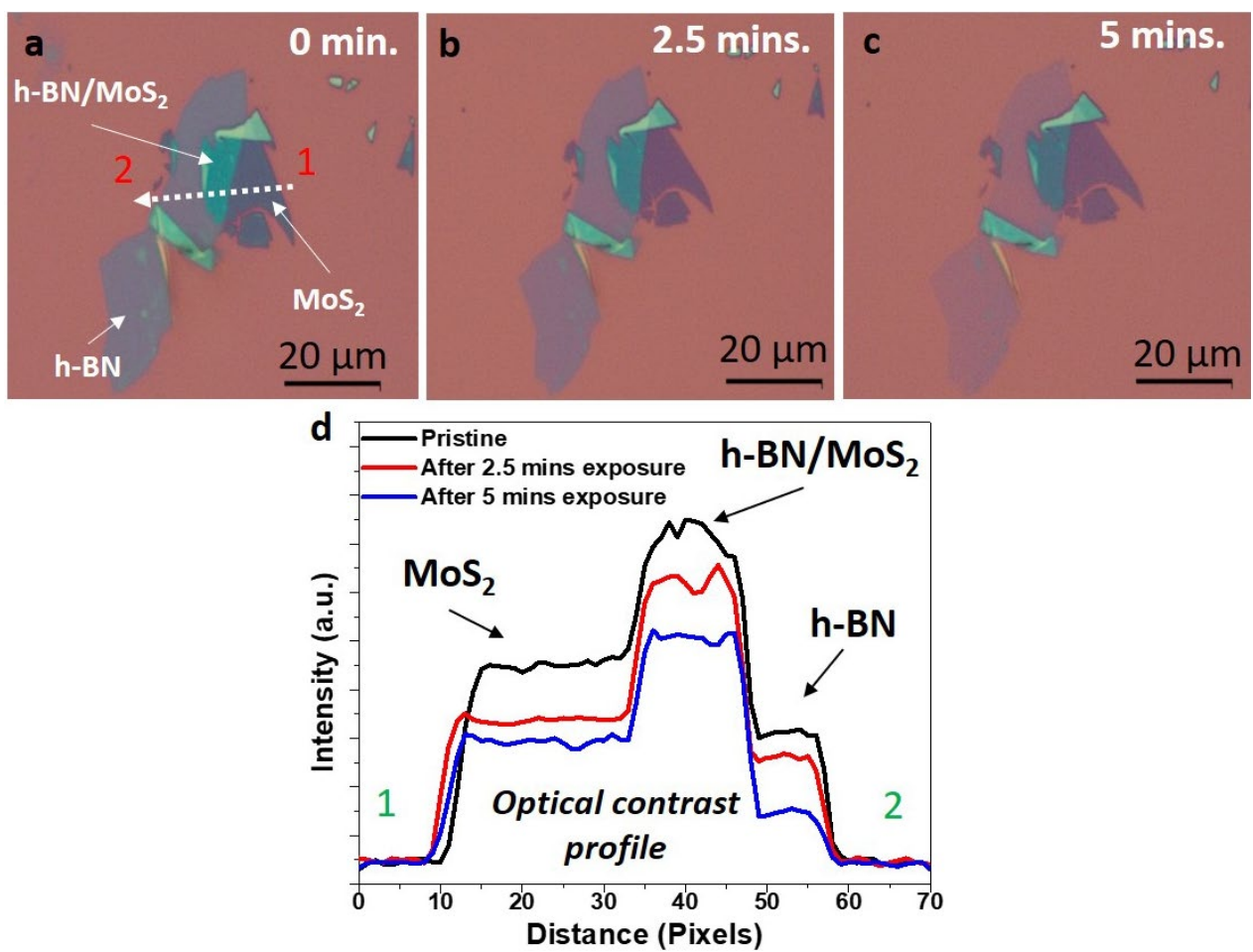

**Fig. S3:** (a-c) Optical microscopy images of a shielded heterostructure (h-BN stacked partly on top of few-layer MoS$_2$) as a function of etching time. (d) line profile of the optical image contrast from the region indicated by the white arrow in (a).

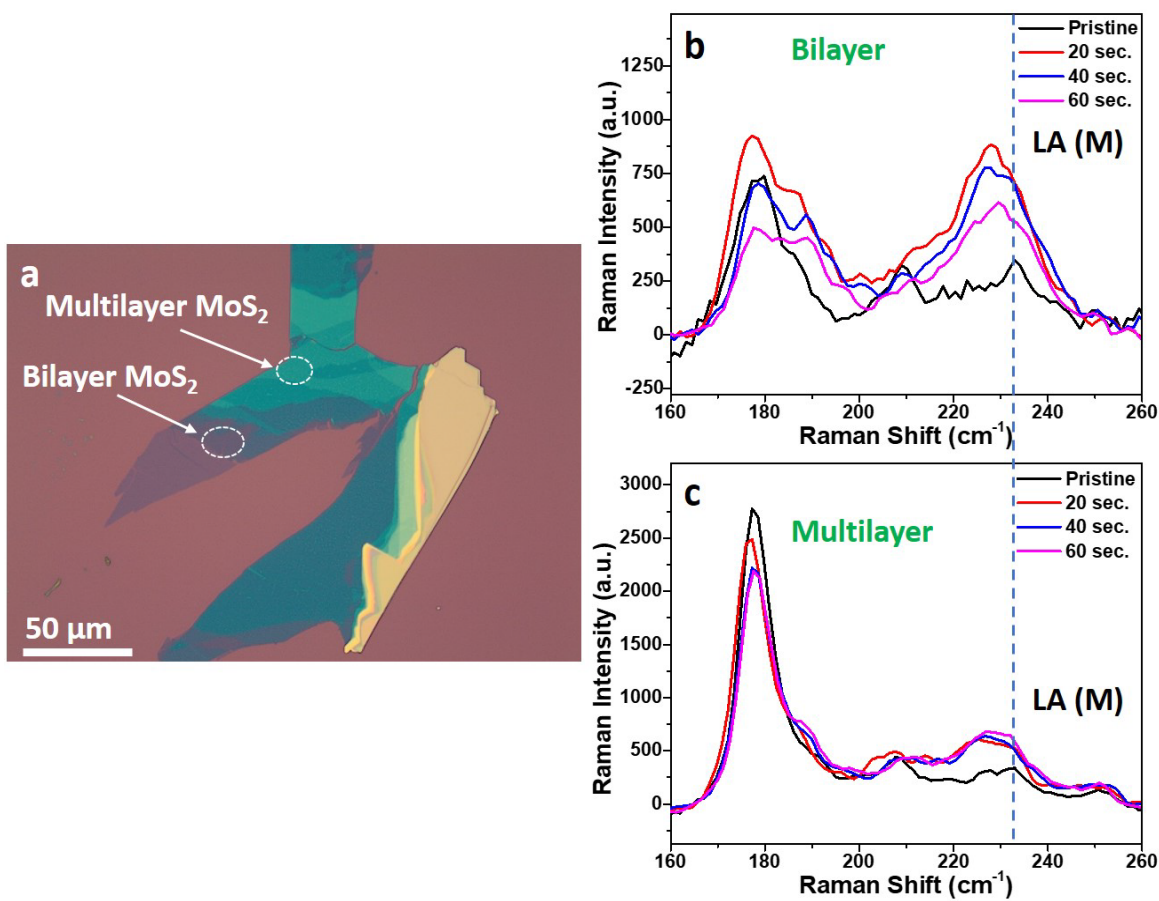

**Fig. S4:** (a) An optical image of monolayer to few-layer MoS$_2$ transferred onto a SiO$_2$/Si substrate; (b) Raman spectra from an unshielded bilayer of MoS$_2$, as a function of exposure time; (c) Raman spectra from an unshielded multilayer of MoS$_2$, as a function of exposure time. Defect formation is apparent in both samples following 20 secs of plasma exposure.

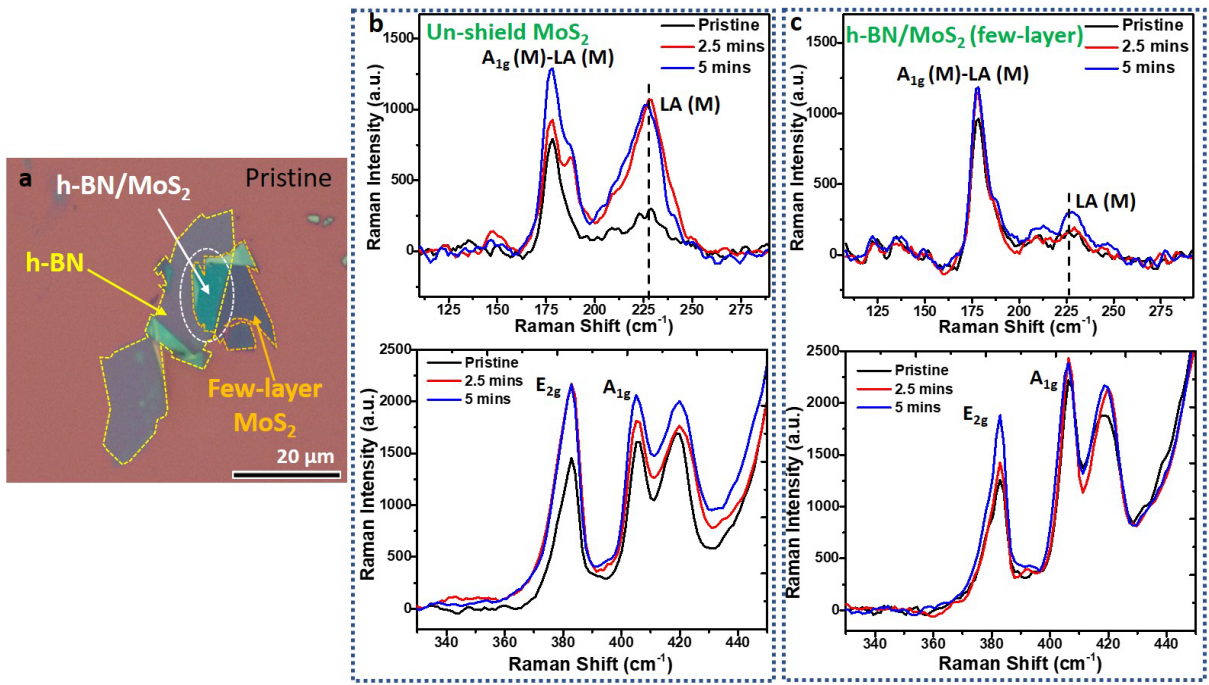

Figure S5: (a) Optical image of few-layer MoS$_2$ in the form of van der Waals heterostructure, h-BN/MoS$_2$ and corresponding defect Raman mode (LA(M)) as well as characteristic Raman modes (E$_{2g}$ and A$_{1g}$) analyzed for the (b) Un-shielded and (c) shielded MoS$_2$ regions.

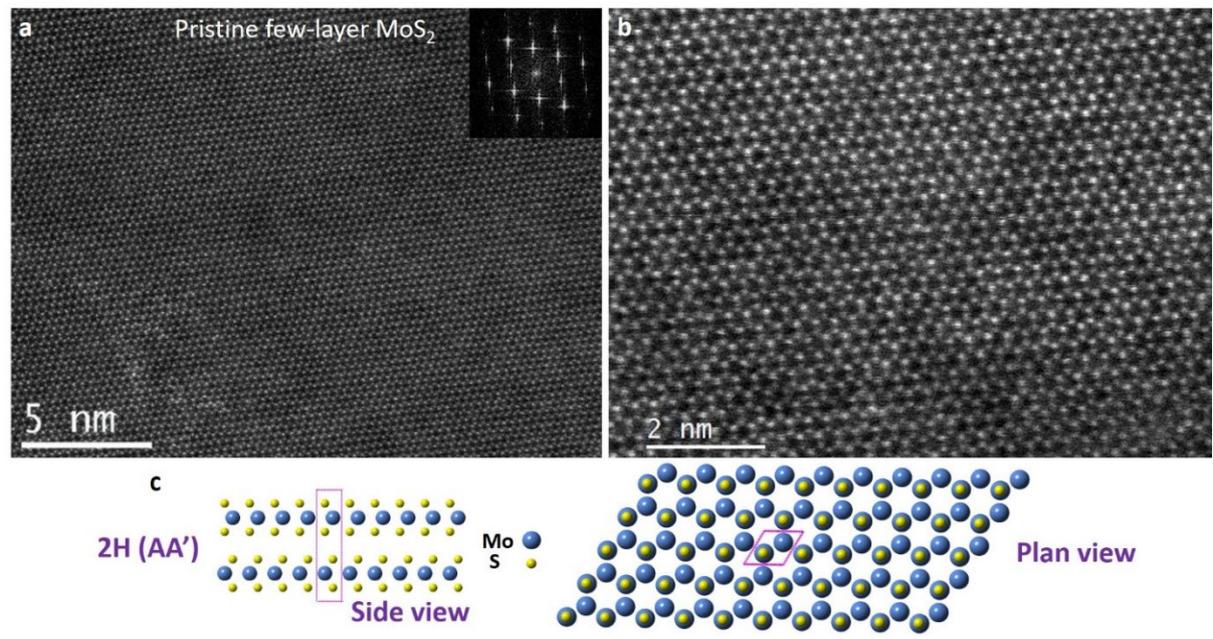

**Fig. S6:** (a-b) Atomic scale HAADF-STEM images of few-layer MoS$_2$ as exfoliated and transferred to the TEM support grid. Inset FFT pattern indicates the high degree of crystallinity of the MoS$_2$ layer. Additionally, the images show that there is no significant defect formation induced by the electron beam in these imaging conditions. (c) Corresponding atomic model of 2H (AA') phase of MoS$_2$ (cross section and plan view).

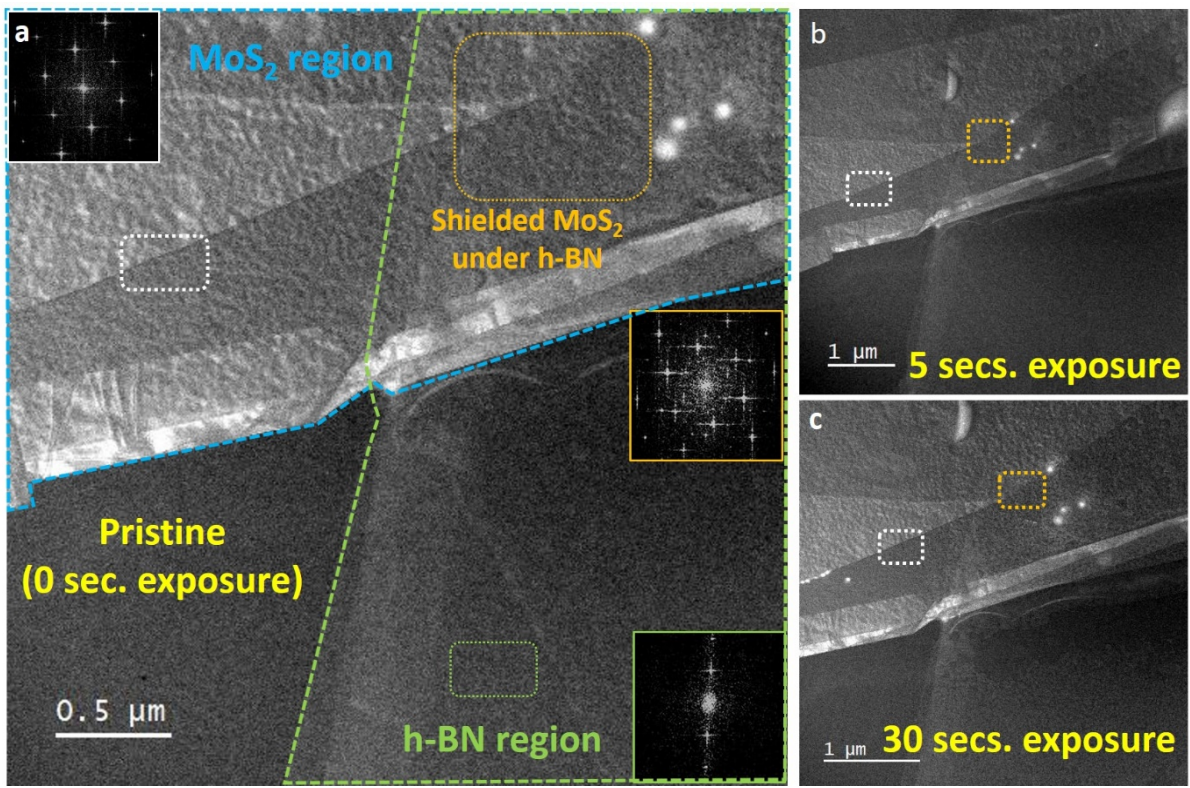

**Fig. S7:** Low magnification ADF-STEM images of h-BN/MoS$_2$ heterostructure. (a) Sample prior to Ar$^+$ plasma exposure, with individual regions indicated (white and orange rectangle for unshielded vs shielded regions respectively). (b) Following 5 seconds of plasma exposure; (c) Following 30 seconds of exposure.

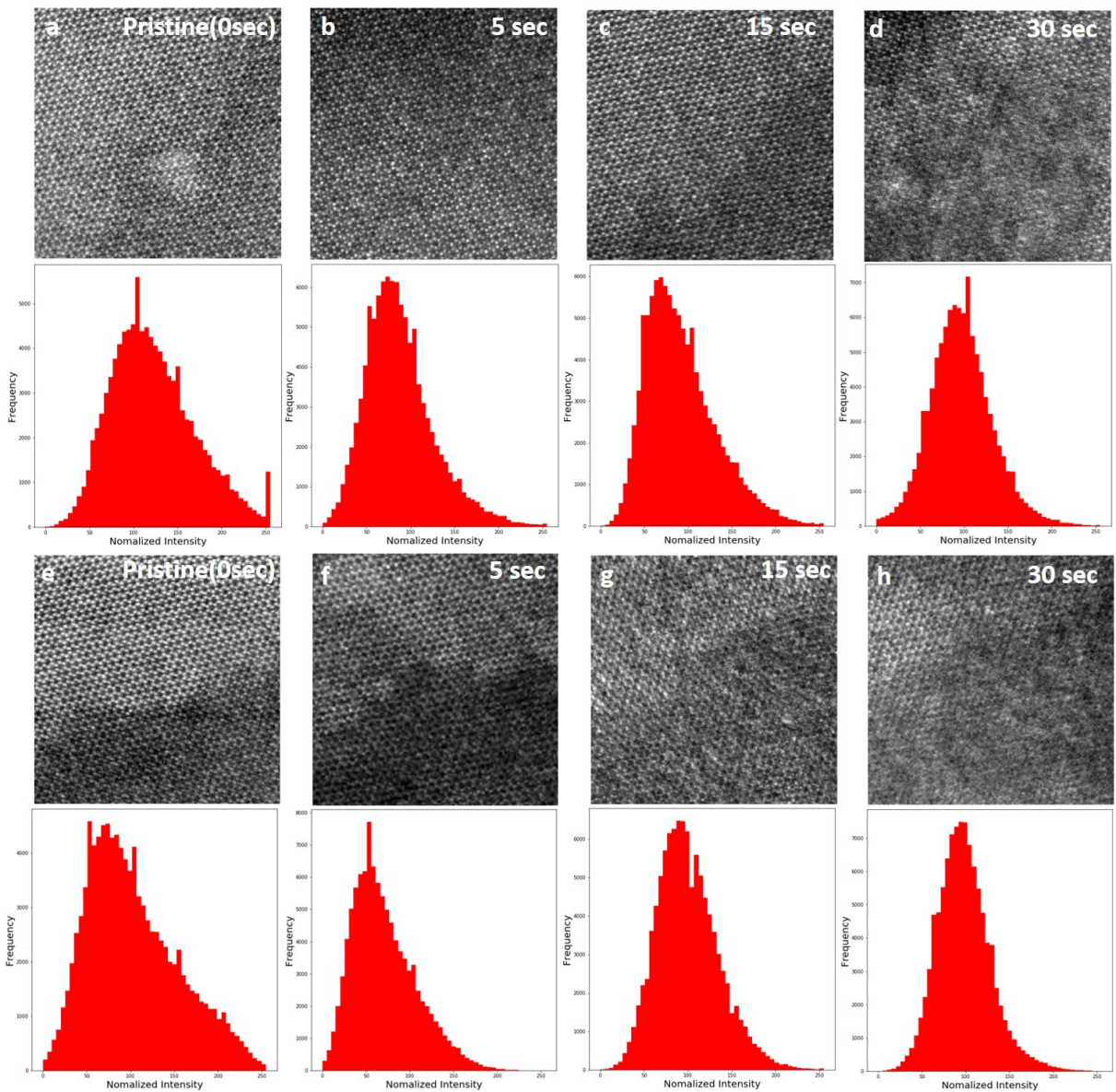

Figure S8: STEM imaging of the (a-d) shielded heterostructure (h-BN/$MoS_2$) as well as (e-h) unshielded $MoS_2$ region at different plasma exposure time along-with corresponding histogram profile normalized by respective image maximum intensity.

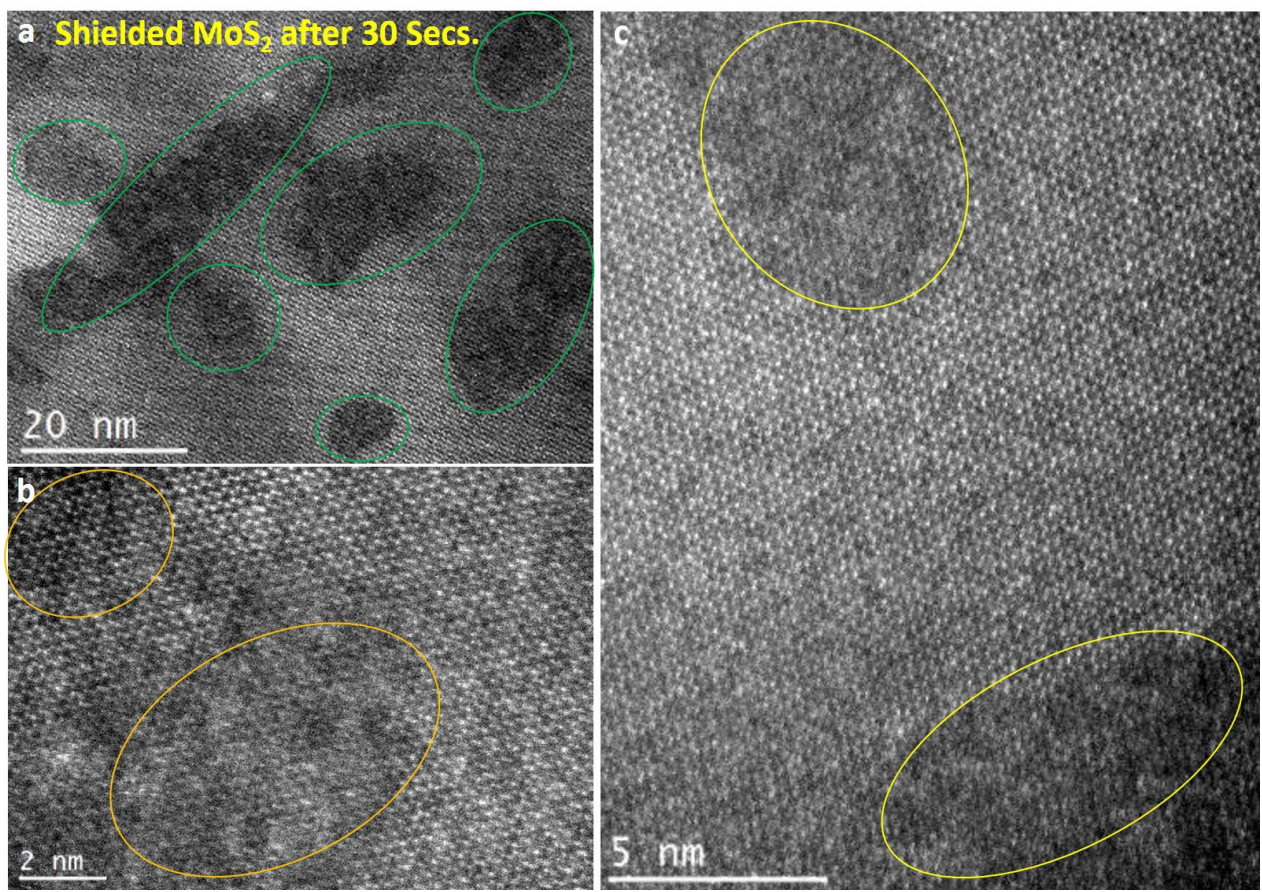

**Fig. S9:** (a-c) Atomic-scale ADF STEM images from the shielded MoS$_2$ after 30 secs of plasma exposure. Defective patches are present in the MoS$_2$ after penetrating the h-BN shielding layer.

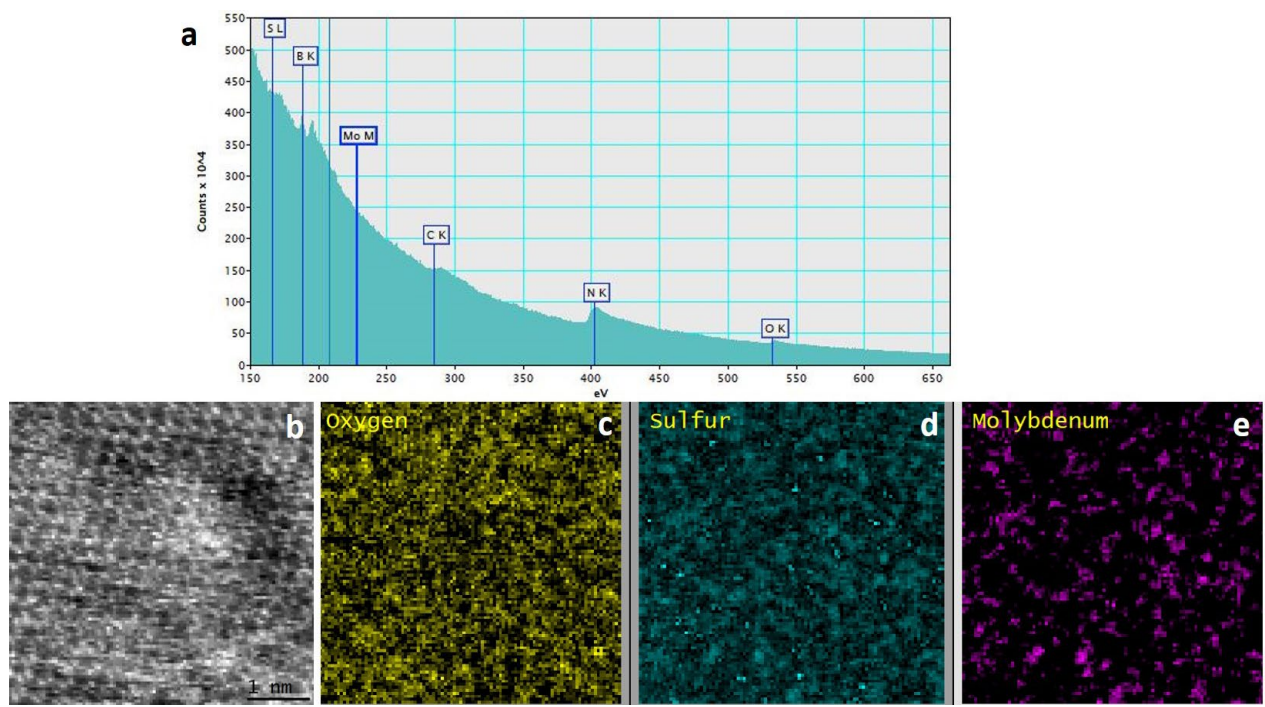

**Fig. S10:** An atomic scale electron energy loss spectroscopy (EELS) analysis from few-layer MoS$_2$ after 15 secs of plasma exposure. (a) EELS spectrum showing the core-loss signatures corresponding to sulfur, boron, molybdenum, carbon, nitrogen and oxygen. (b-e) ADF-STEM image and EELS elemental maps.

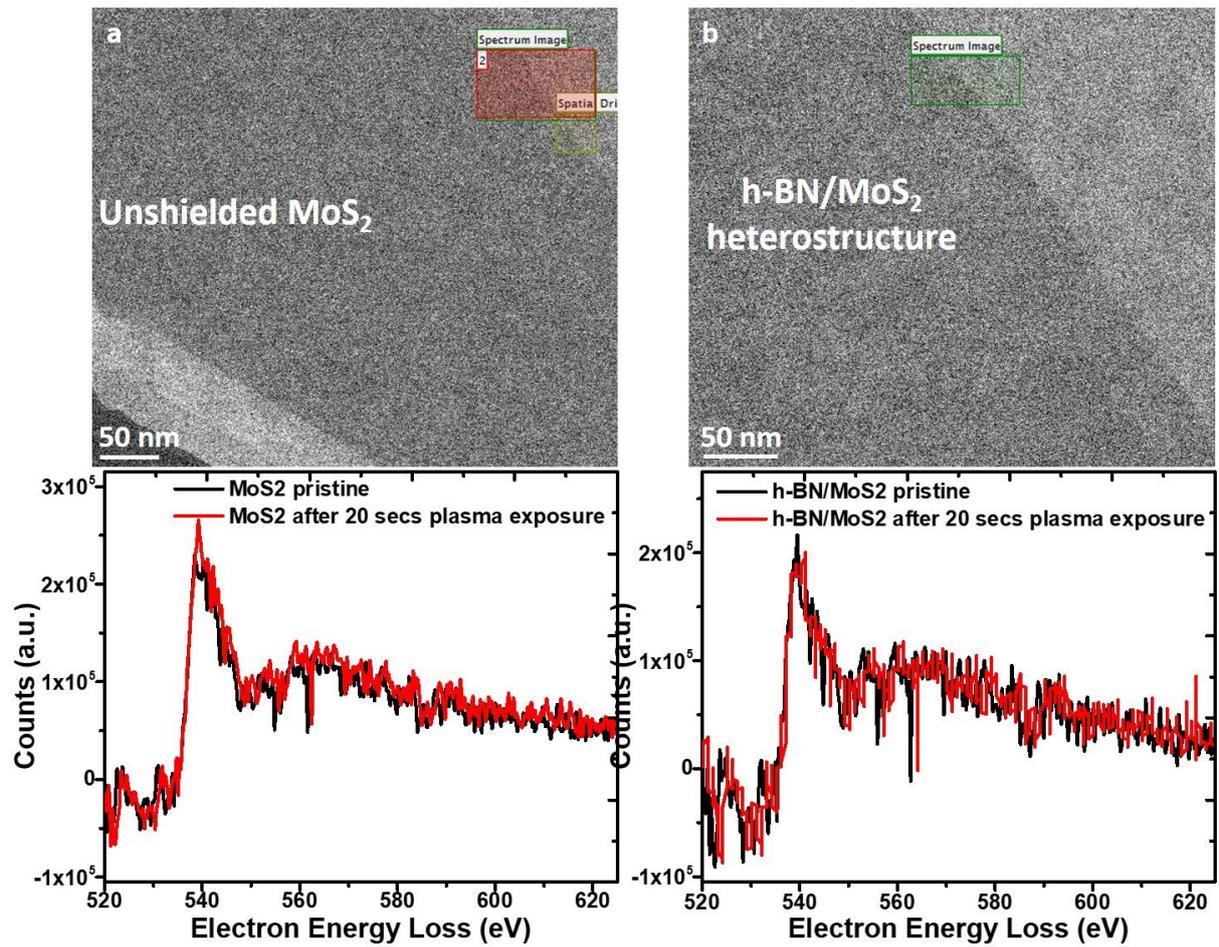

Figure S11: Oxygen content analysis from core-loss EELS spectra for (a) unshielded and (b) shielded MoS$_2$ layer before and after 20 secs of continuous plasma exposure time.

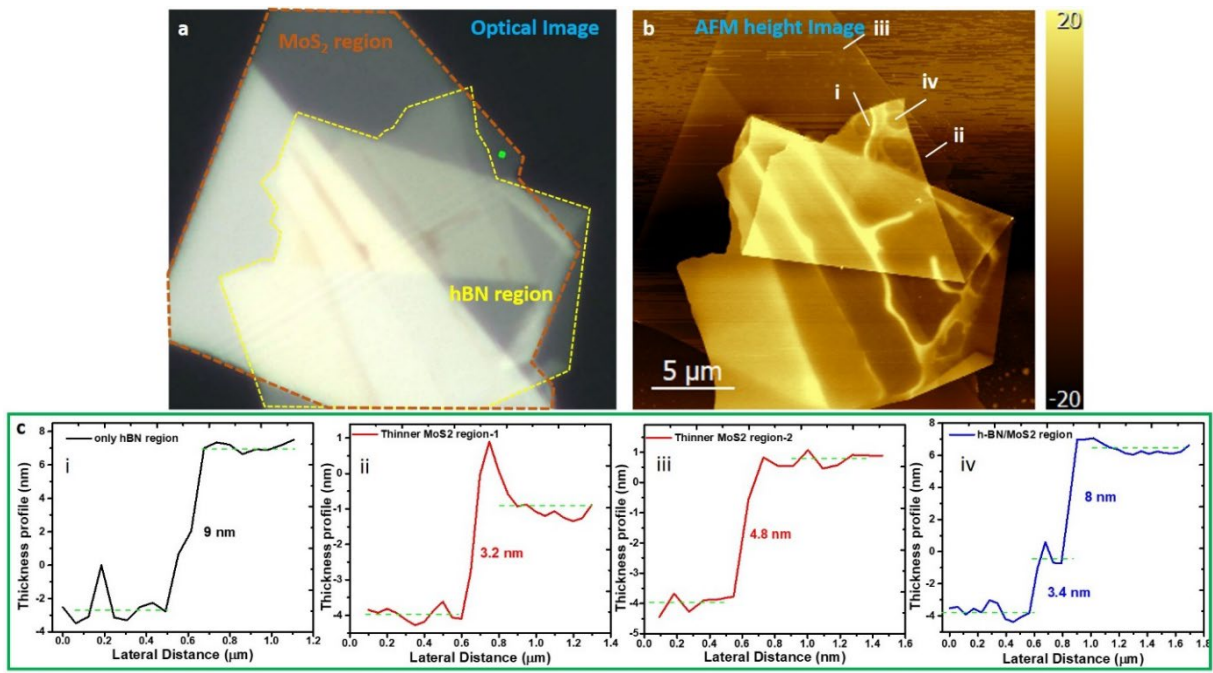

Figure S12: (a) Optical micrograph of the h-BN/MoS$_2$ heterostructure where orange marked region shows MoS$_2$ flake while yellow marked region for h-BN flake. (b) AFM height image and corresponding (c) thickness profile for different portion of the h-BN, MoS$_2$ as well as heterostructure regions.

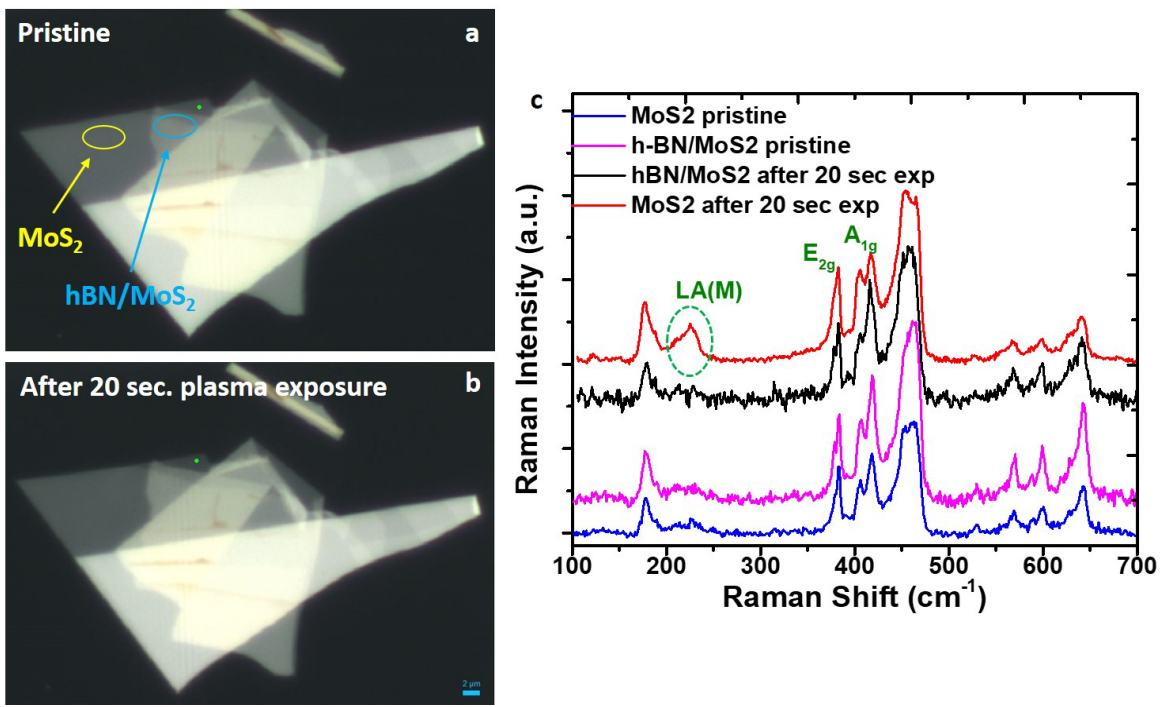

Figure S13: Optical micrograph of the h-BN/MoS$_2$ heterostructure (a) before and (b) after the plasma exposure of 20 secs (utilizing dedicated TEM plasma cleaner system). Corresponding Raman modes analysis across shielded (h-BN/MoS$_2$) and unshielded (MoS$_2$) regions, as marked in (a) of the heterostructure sample. A clear evolution of defect mode (LA(M) ~226 cm$^{-1}$) in the unshielded MoS$_2$ after 20 secs of plasma exposure.

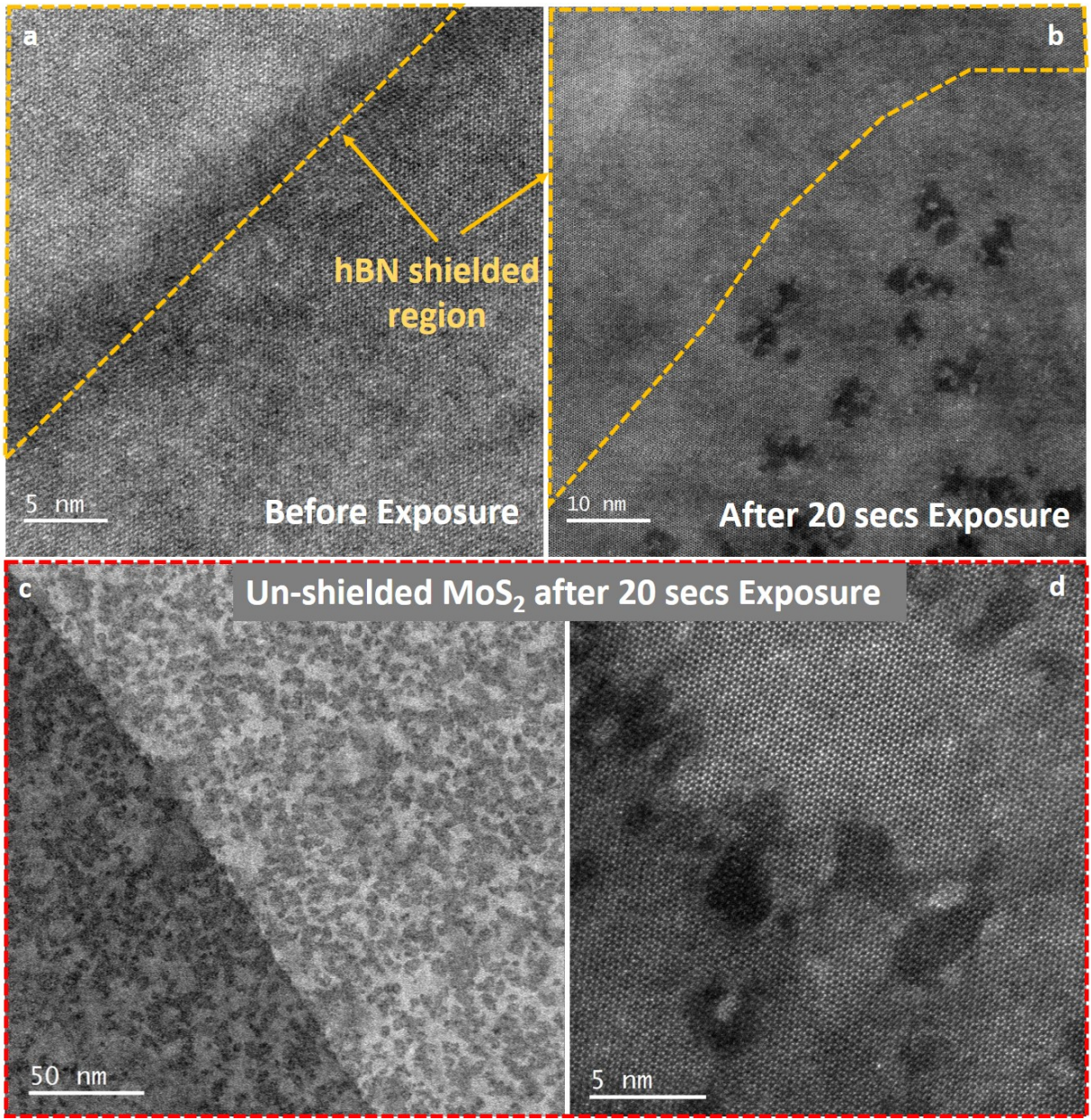

Figure S14: STEM imaging before and after continuous 20 secs plasma exposure for the (a, b) shielded (hBN/$MoS_2$) and (c, d) unshielded $MoS_2$ regions.

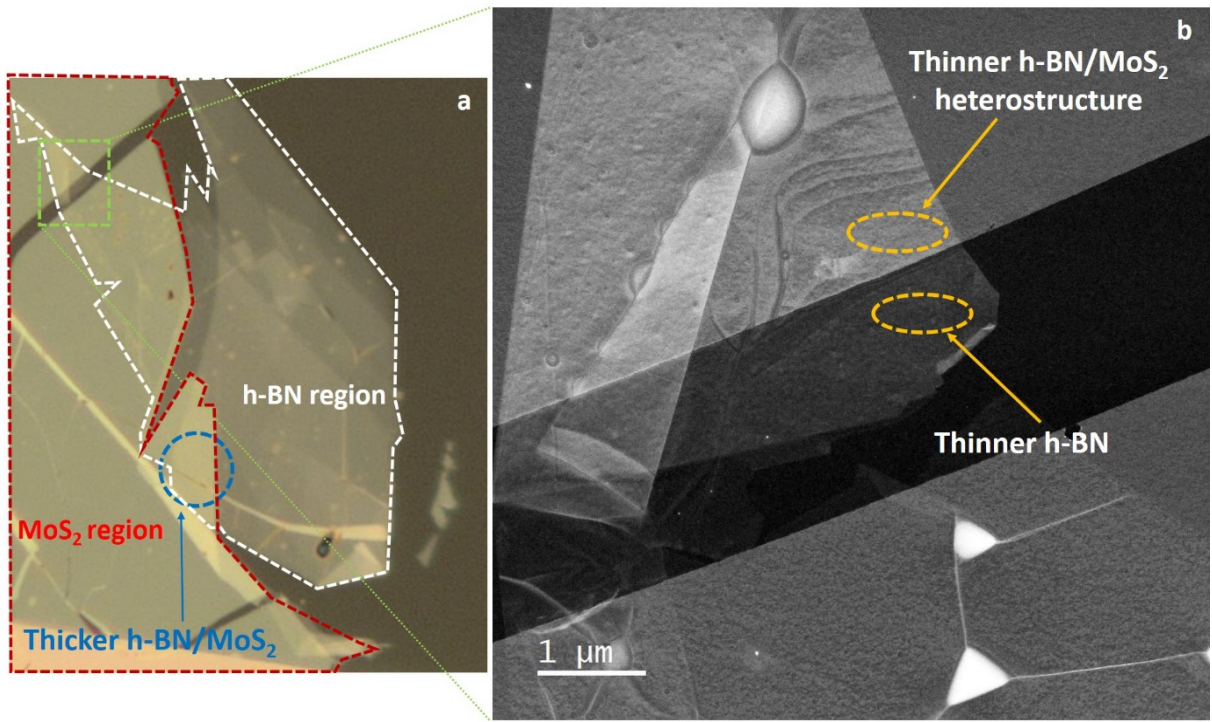

**Fig. S15:** (a) Optical image showing h-BN and $MoS_2$ layers, as indicated. (b) Corresponding ADF-STEM images from the region indicated in (a).

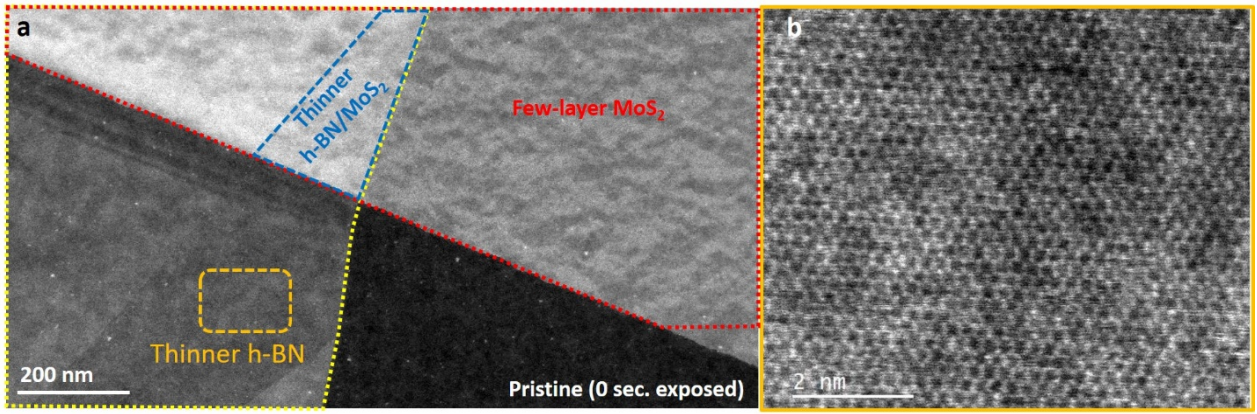

**Fig. S16:** (a) Low magnification and (b-d) atomic resolution HAADF-STEM images of the thinner h-BN layer prior to plasma exposure, from different locations.

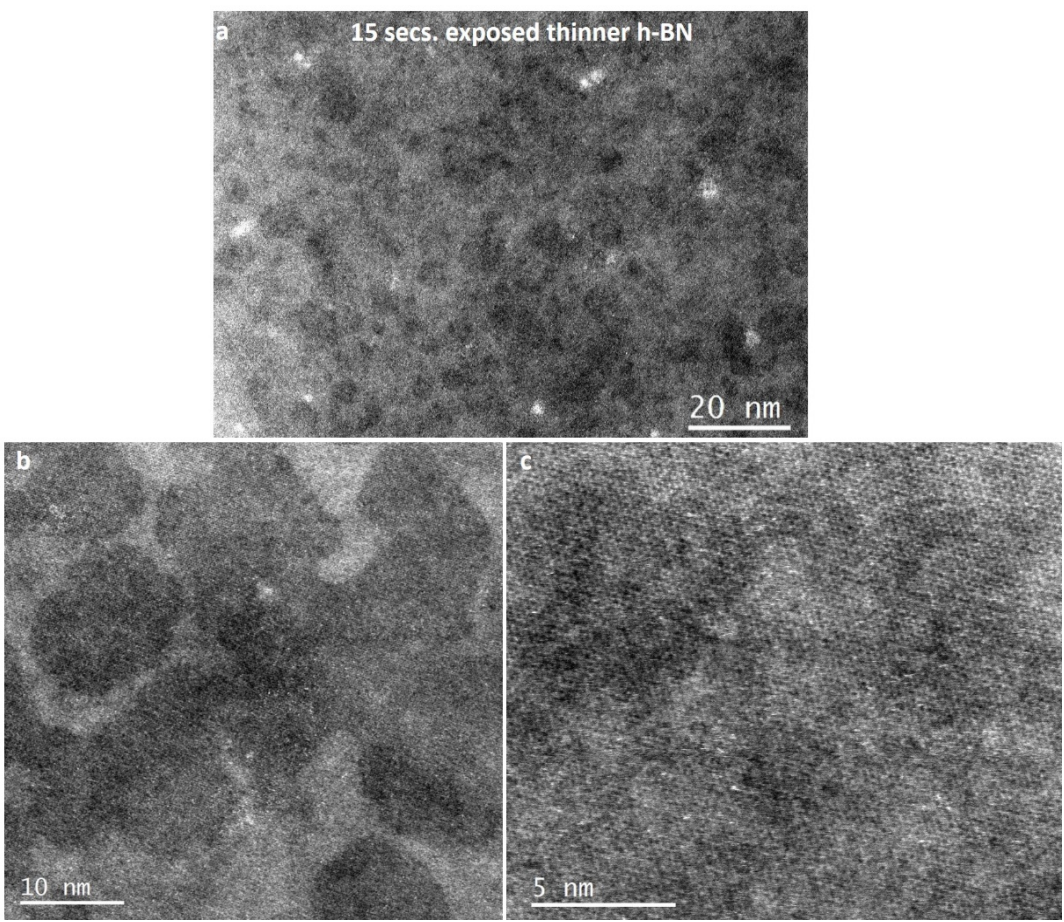

**Fig. S17:** (a-b) Low magnification and (c) atomic resolution HAADF-STEM imaging of the thinner h-BN layer after 15 secs continuous plasma exposure. Plasma damaged regions are visible as dark contrast.

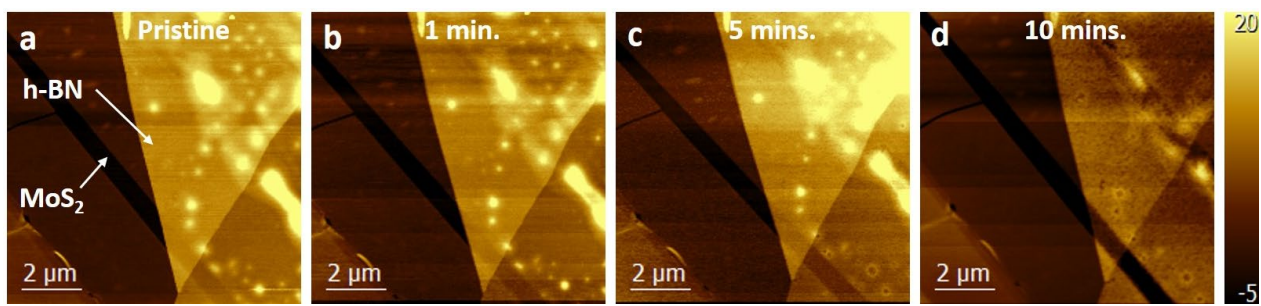

**Fig. S18:** AFM images of h-BN/MoS$_2$ heterolayers as a function of different plasma exposure time (0-10 mins). (a) Heterostructure of h-BN and MoS$_2$ in its pristine condition and then following (b-d) different plasma exposure. The images show regions of defect clustering.